\documentclass[prd,twocolumn,preprintnumbers,superscriptaddress]{revtex4}
\usepackage{epsfig}

\newcommand{\beq}{\begin{equation}}
\newcommand{\eeq}{\end{equation}}

\begin{document}

\title{Monopole clusters at short and large distances}

\author{V.G. Bornyakov}
\email{bornvit@sirius.ihep.su} 
\affiliation{Institute for High Energy Physics, Protvino
142284, Russia}

\author{P.Yu. Boyko}
\email{boyko@itep.ru} 

\author{M.I. Polikarpov}
\email{polykarp@heron.itep.ru}
\affiliation{ Institute of Theoretical and  Experimental
Physics, B.~Cheremushkinskaya~25, Moscow, 117259, Russia}

\author{V.I. Zakharov}
\email{xxz@mppmu.mpg.de}
\affiliation{Max-Planck Institut f\"ur Physik, F\"ohringer Ring 6, 80805,
M\"unchen, Germany}

\preprint{ITEP-LAT/2003-05}
\preprint{MPI-PhT 2003-18}

\begin{abstract}
We present measurements of various geometrical characteristics of monopole
clusters in $SU(2)$ lattice gauge theory. The maximal Abelian projection is
employed and both infinite, or percolating cluster and finite clusters are
considered. In particular, we observe scaling for average length of segments of
the percolating cluster between self-crossings, correlators of vacuum monopole
currents, angular correlation between links along trajectories. Short clusters
are random walks and their spectrum in length corresponds to free particles. At
the hadronic scale, on the other hand, the monopole trajectories are no longer
random walks. Moreover, we argue that the data on the density of finite
clusters suggest that there are long-range correlations between finite clusters
which can be understood as association of the clusters with two-dimensional
surfaces, whose area scales.
\end{abstract}

\maketitle
\section{Introduction}

The interest in monopoles in non-Abelian gauge theories
is mostly due to the dual superconductor mechanism of the confinement,
for review see \cite{reviews}. The mechanism assumes condensation of
magnetic monopoles in the vacuum.
The idea is supported by the lattice data and
the phenomenology of the lattice monopoles is quite rich.
Especially, in case of $SU(2)$ gluodynamics which we will concentrate on
in this paper.

Detailed theoretical interpretations of the data appear, however, difficult
because monopoles are defined not directly in terms of the original $SU(2)$
fields but rather in terms of projected fields. The use of  a projection is
rooted in the fact that monopoles are intrinsically $U(1)$ objects and there
are infinitely many  ways to select a $U(1)$ subgroup for the monopole
definition. In particular, monopoles of the maximal Abelian projection (MAP)
are defined in the following sequence of steps (which we describe in a somewhat
simplified way). First, one finds the maximal Abelian gauge maximizing the
functional:
\beq 
\label{mag} R_{Abel} = \int d^4 x~ \left[\left( A^1_\mu(x)\right)^2 + \left(A^2_\mu(x)\right)^2 \right]
\eeq
where $A^a_\mu(x)$ are components of the gauge field.
Finally, monopoles are defined as singularities of the ${A^3_\mu(x)}$ fields
which violate the Bianchi identities: \beq\label{bianchi}
j^{mon}_{\mu}(x)~=~\epsilon^{\mu\nu\rho\sigma}\partial_{\nu}\partial_{\rho}
{A}^3_{\sigma}(x)~~,
\eeq
where $j_\mu^{mon}(x)$ is the monopole current and all the expressions
are well defined on the lattice.

Geometrically, monopoles are represented by closed trajectories
(on the dual lattice) and one usually discusses properties of the monopole
clusters. In particular, it is important to distinguish between
the percolating cluster and finite clusters. The percolating cluster
fills in the whole of the lattice and is in a single copy for each field
configuration. Finite clusters are characterized, in particular,
by their spectrum as function of the length $l$. Commonly, one introduces
the corresponding densities of the percolating and finite clusters:
\beq\label{l}
l_{perc}~\equiv~4~\rho_{perc} a^4 N_{sites},~~l_{fin}~\equiv~4~\rho_{fin} a^4 N_{sites}~~,
\eeq
where $l_{perc},l_{fin}$ is the total length of the corresponding clusters,
$N_{sites}$ is the number of lattice sites and $a$ is the lattice spacing.

An important question is whether position of the monopoles
is of any physical significance. On one hand, monopole
trajectories are distinguished by singularities of the projected fields,
see (\ref{bianchi}), and they are point like in terms of the projected fields.
On the other hand, one can suspect that these singularities are
artifacts of the projection.

Probably, a priori the latter possibility looks more reasonable since the
monopole definition involves a non-local gauge fixation, see (\ref{mag}).
However, there are accumulating lattice data which indicate that in the maximal
Abelian gauge the monopoles, which by construction have size of the lattice
spacing $a$, are physical. In particular, the density of the percolating
monopoles scales: \beq \label{rho1} \rho_{perc}=0.65(2)~\sigma_{SU(2)}^{3/2}\, ,
\eeq
where $\sigma_{SU(2)}$ is the string tension~\footnote{In estimates we will
also use $\Lambda_{QCD}$ although we study actually $SU(2)$ gluodynamics,
$\sigma_{SU(2)}\sim \Lambda^2_{QCD}$.} and we quoted the data from
Ref.~\cite{mueller} where references to earlier papers can also be found.

Let us emphasize that the result (\ref{rho1}) implies that the corresponding 
monopole trajectories are meaningful even on the scale $a$. 
Indeed, the total length of the percolating clusters
which scales in the physical units is added up from small steps of size $a$ and
finally is independent on $a$. Thus, one could argue, see, e.g., \cite{hart},
that there are gauge invariant objects which are detected through the
projection. It is of course an intriguing hypothesis which is worth to be
thoroughly checked. Unfortunately, this suggestion cannot be scrutinized
theoretically since the anatomy of the lattice monopoles in terms of the
non-Abelian fields is largely unknown, for review see, e.g., \cite{coleman}.
Rather one should apply further phenomenological tests.

In particular, one can measure  non-Abelian action associated
with the monopoles, see \cite{anatomy} and references
therein. For monopoles in the percolating clusters it turns out that, at least 
at presently available lattices, the action corresponds to a monopole mass 
which diverges in the ultraviolet:
\beq\label{action}
M(a)~\sim~{const\over a}~~.
\eeq
$M(a)$ is the lattice analog of the `magnetic mass' in the continuum,
$M(a)~\sim~\int {\bf H}^2d^3r$ where ${\bf H}$ is the magnetic field.
The result (\ref{action}) strongly suggests that the lattice monopoles 
are associated with singular non-Abelian fields and  the singularity
(\ref{bianchi})
is {\it not} an artifact of the projection.

The next question is how singular fields can be important at all since their
contribution  in the limit $a\to 0$ is suppressed by an infinite
action~\footnote{Hereafter we tacitly assume that the $a$ dependence observed
at the presently available lattices will persist at smaller lattice spacings as
well. The smallest $a$ available now is about $a\approx (3~GeV)^{-1}$.}. An
apparent answer to this question is that the action and entropy factor are both
divergent in the ultraviolet but cancel each other in such a way that, say, the
total density of monopoles in the percolating cluster scales in the physical units, see (\ref{rho1}). An example of such
a fine tuning is provided by  $U(1)$ theory, see, in particular,
\cite{polyakov}. In that case the fine tuning is ensured by choosing tuned
values of the electric charge. In case of non-Abelian theories the hypothesis
on the fine tuning \cite{vz,maxim} is rather phenomenological and the mechanism
of the fine tuning is to be understood yet.

To summarize,

(1) there is evidence that at presently available lattices the monopoles
defined within the maximal Abelian projection are associated in fact with
singular non-Abelian fields fine tuned to the corresponding entropy factors.

(2) This means that the structure of the non-perturbative  vacuum fluctuations
is quite different from the standard picture of `bulky' fields of the size
$\Lambda_{QCD}^{-1}$.

(3) Although the conclusion on the fine tuning is strongly suggested by the
data like (\ref{rho1}), (\ref{action}) further checks of it are highly
desirable. Indeed, the whole basis is the phenomenology of the lattice
monopoles and further data can bring further insight.

Motivated by these considerations, we have investigated in more detail the
scaling properties of the monopole clusters~\footnote{The preliminary numerical
results were presented at the ``Lattice 02'' conference \cite{latt02}.}. The
outline of the paper is as follows. In Sect.~\ref{sect2} we present detailed
studies of the geometrical elements of the percolating cluster. Namely, the
cluster consists of self-crossings and segments between the crossings. We
report on the measurements of distribution in the length of the segments,
long-range correlations between directions of the links occupied by
monopole currents, scaling properties of the segments characteristics. In
Sect.~\ref{clusters_sect} we first present data on the spectrum of the finite
size clusters (subsection~\ref{short}) and then on their density
(subsection~\ref{densities}). In subsection~\ref{sub33} we present results on
the correlations of the monopole trajectories at large distances which are
sensitive, at least in principle, to the glueball masses. Discussions of the
observations made in this work are in Sect.~\ref{discussion}. Finally, some details of
simulations are given in the Appendix.


\section{Segments of the percolating cluster}
\label{sect2}
\subsection{Distribution in the segments length}

The percolating cluster consists of segments (that is, trajectories between
crossings) and crossings. The segments, in turn,  are made from the
links on the dual lattice. We will consider
the length of the segment, $l_{segm}$  and the Euclidean distance between
the end points of the segments, $d$.

\begin{figure}
\begin{center}
\includegraphics[width=\columnwidth]{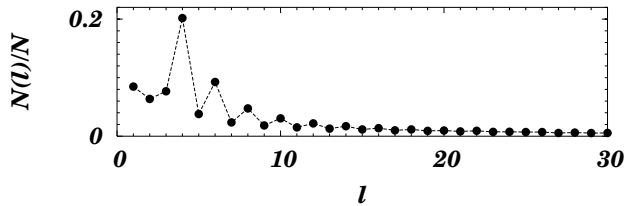}
\caption{The distribution $N(l)/N$, $\beta = 2.4$, the lattice size is $24^4$.}
\label{lsegment_fig}
\end{center}
\end{figure}

In Fig.~\ref{lsegment_fig} we show the normalized distribution, $N(l)/N$, of
the lengths of the segments; $N(l)$ is the number of the segments with the
length $l$ in lattice units, $N$ is the total number of segments.

\begin{figure}
\begin{center}
\includegraphics[width=\columnwidth]{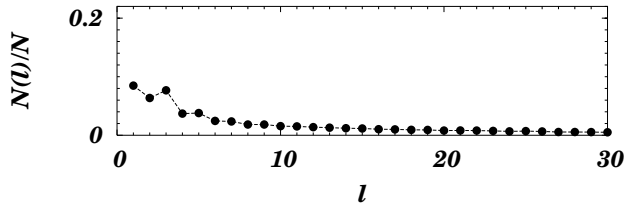}
\caption {$N(l)/N$ for trajectories which are not closed loops; $\beta$ value
and the lattice size are the same as in Fig.\ref{lsegment_fig}}
\label{lsegmentnc_fig}
\end{center}
\end{figure}

The oscillations of $N(l)/N$ are clearly seen, the number of trajectories with
the even $l$ is systematically larger than that with the odd $l$. This effect
is due to the large number of the closed trajectories (loops), which are
segments with $d=0$ and $l \neq 0$. In Fig.~\ref{lsegmentnc_fig} we show
$N(l)/N$ only for not closed trajectories, $d \neq 0$.

The oscillations are substantially reduced compared with
Fig.~\ref{lsegment_fig}.

\subsection{Long-range correlation between links directions}
\label{linksdirection}

\begin{figure}
\begin{center}
\includegraphics[width=\columnwidth]{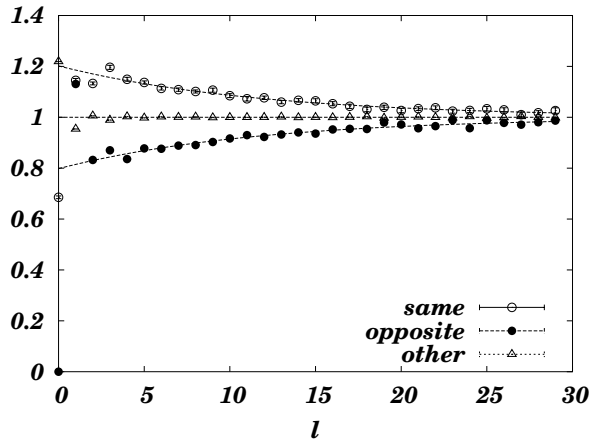}
\caption{Correlations of the directions of the links lying on the monopole
current; $\beta=2.6$, the lattice is $28^4$.} \label{memory_fig}
\end{center}
\end{figure}

Next, we turn to angular correlation between the links on the monopole
trajectory. Let us call link $C_0$
belonging to the percolating cluster as the "initial" one. Then one can measure
the probability that the link $C_l$, connected to the link $C_0$ by the 
monopole trajectory of length  $l$,  
has the same direction. Of course this probability is a decreasing
function of $l$, but it occurs that there exists a "long memory" of the initial
direction. This fact is illustrated in Fig.~\ref{memory_fig} where the
correlation of the direction of the initial link $C_0$ with the direction of
the link $C_l$ is shown. The direction of $C_l$ can be the same as that of 
$C_0$, or
opposite, or else (neither the same nor the opposite). We  normalize all
three probabilities in such a way that they are equal to unity for random walk if $l\neq 0$ (for $l=0$
the opposite direction is forbidden).
From Fig.~\ref{memory_fig} it is seen that even for links separated by 17
lattice steps the most probable direction is the same as the direction of the
initial link. For smaller values of $\beta$ the correlations of the directions
exist for not so large values of $l$. The deviation from unity of the 
probability to have the same direction falls off  exponentially: 
$P_{same} = 1 + A_s e^{-\mu_s la}$. The
probability to have the opposite direction behaves as follows: 
$P_{opposit} = 1 - A_o
e^{-\mu_o la}$. These fits are shown by dashed lines in Fig.~\ref{memory_fig}.
The masses $\mu_s$ and $\mu_o$ coincide within the error bars and are
independent of the lattice spacing, see Table~\ref{mu}.
\begin{table}[!h]
\begin{center}
\caption{The fitted masses
from data for $P_{same} - 1$ and $1 - P_{opposit}$}
\begin{ruledtabular}
\begin{tabular}{c l l l l}
$\beta$ & 2.45 & 2.50 & 2.55 & 2.60 \\
$ \mu_{s}$, MeV & 290(60) & 290(20) & 271(15) & 273(12) \\
$\mu_{o}$, MeV & 250(60) & 240(20) & 252(15) & 277(15) \\
\end{tabular}
\end{ruledtabular}
\label{mu}
\end{center}
\end{table}
\subsection{Scaling properties of the segments}
\label{ld_sec}

It was already mentioned in the Introduction that the density of the
percolating cluster scales, that is $\rho_{perc}$ has a well defined value in
the continuum. Here we will present further data on the scaling properties of
geometrical elements of the percolating cluster.

\begin{figure}
\begin{center}
\includegraphics[width=\columnwidth]{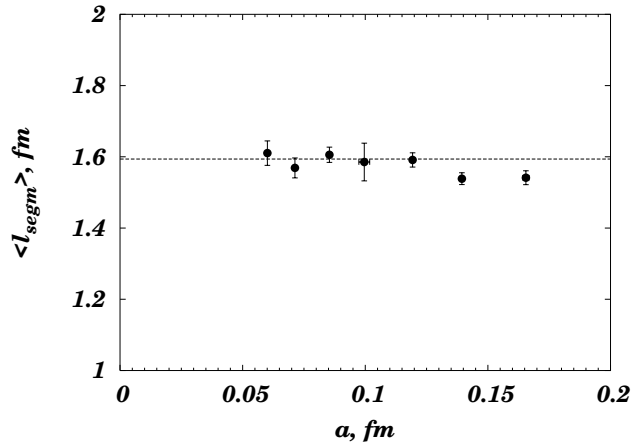}
\caption {Average segment length, $\langle l_{segm}\rangle$, vs. lattice spacing.}
\label{lscaling_fig}
\end{center}
\end{figure}

First, we observe that the average length of the segment of the monopole 
trajectory
between crossings, $\langle l_{segm}\rangle$, does not depend
on the lattice spacing, see Fig.~\ref{lscaling_fig}. Making fit by a constant 
for $\beta > 2.35$ we obtain

$$\langle l_{segm} \rangle~ =~ 1.60(1)~fm.$$
The way used to convert lattice results into physical units
is explained in Appendix. The lattice spacing values $a$ for various  
$\beta$'s are  also given in Appendix, see Table III.

\begin{figure}
\begin{center}
\includegraphics[width=\columnwidth]{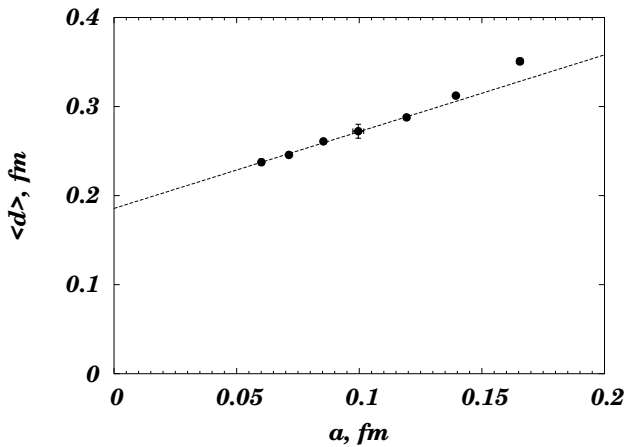}
\caption {$\langle d \rangle $ vs. lattice spacing. The dashed line shows the linear fit for 
$\beta > 2.35$.} 
\label{dscaling_fig}
\end{center}
\end{figure}

In the average Euclidean distance  between the crossings $\langle d\rangle$  the
violations of the scaling are more significant, see Fig. 5. However, the
deviations from the scaling can be approximated for $\beta > 2.35$ by linear in $a$ corrections.
Then in the continuum limit $a \to 0$, $\langle d\rangle$ has a non-vanishing value
($\approx 0.20\, fm$), as it is seen from Fig.~\ref{dscaling_fig}.

\begin{figure}
\begin{center}
\includegraphics[width=\columnwidth]{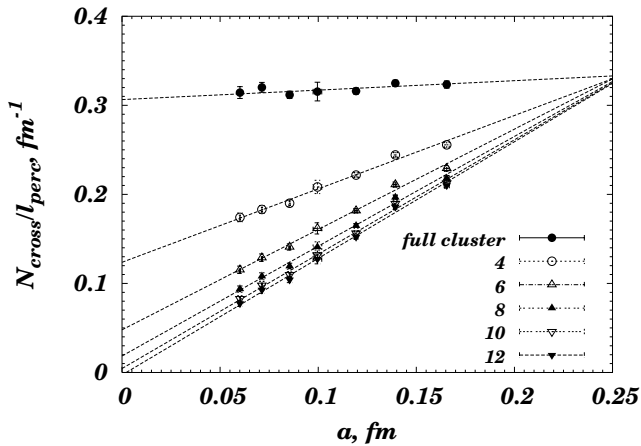}
\caption {Number of crossings per unit physical length vs. lattice spacing in
full percolating cluster and in the percolating cluster with excluded closed
loops. Numbers $4, 6, ... 12$ in the legend mean that closed loops of the length up to $4, 6, ... 12$ are
excluded. The dashed lines show the linear fits.} \label{Ncrossings_fig}
\end{center}
\end{figure}

In Fig.~\ref{Ncrossings_fig} the number of crossing points per unit length of
the monopole trajectory is shown by open circles. This number weakly depends on
$a$ and seems to have the continuum limit ($a \to 0$) $\approx 0.3\, fm^{-1}$.

The situation changes if we exclude the closed loops of finite lattice length
connected to the percolating cluster. The number of crossings reduces and is
compatible with zero in the continuum limit if we exclude the loops of the
length  up to 8 (or up to larger length). This fact is illustrated in
Fig.~\ref{Ncrossings_fig}.

The continuum limit value $\langle N_{cross}\rangle/l_{perc} = 0.3 \, fm^{-1}$ for the full
percolating cluster together with the data for the percolating
cluster density (Fig.~\ref{rho_fig}) corresponds to approximately 10 crossings
in percolating cluster per hypercube $1 \,\, fm^4$. To complete the picture of
the structure of the percolating monopole cluster in the continuum limit the 
data for
$\langle l_{segm}\rangle$, $\langle d \rangle$ and 
$\langle N_{cross}/l_{perc}\rangle$ can be compared with the average monopole radius
$\langle\rho_m\rangle \approx 0.05 \,\, fm$ and average inter-monopole distance $\langle R\rangle 
\approx 0.5 \,\, fm$ \cite{anatomy}.

\section{Finite clusters}
\label{clusters_sect}
\subsection{Length spectrum of finite clusters}
\label{short}

\begin{figure}
\begin{center}
\includegraphics[width=\columnwidth]{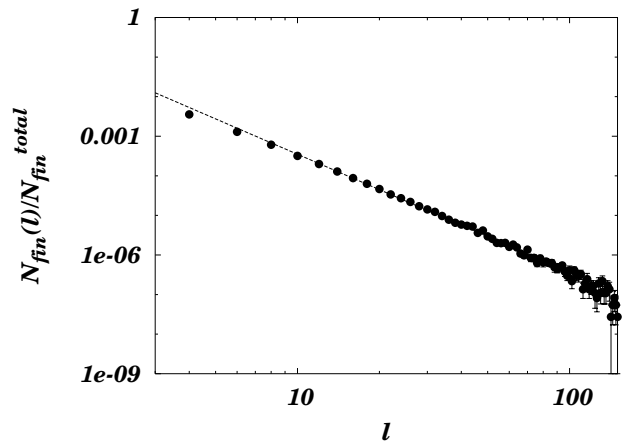}
\caption{The length distribution of finite clusters. $\beta=2.4$, the size of
the lattice is $32^4$.} \label{NUV}
\end{center}
\end{figure}

In Fig.~\ref{NUV} we show the number of the finite clusters of the length $l$
(in the lattice units) vs. $l$. The dashed line in this figure is the fit of the
data by function $Const/l^\alpha$.
It is important that
$1/l^3$ behavior seems to be valid for all considered values of the lattice 
spacings as it can be 
seen from Table~\ref{tbl2} where we show results of the fit for
various $\beta$ values.
As we discuss in Section~\ref{finclust} the dependence $1/l^3$ means that
finite clusters develop in the four-dimensional space. 

\begin{table*}
\begin{ruledtabular}
\begin{tabular}{ c l l l l l l l l l }
$\beta$ & 2.30 & 2.35 & 2.40 & 2.40 & 2.40 & 2.45 & 2.50 & 2.55 & 2.6 \\
L & 16 & 16 & 16 & 24  & 32 &  24 & 24 & 28 & 28 \\
$\alpha$ & 3.12(4) & 3.10(4) & 2.98(2) & 2.95(2)  & 2.970(16) & 2.91(3) & 3.02(3) & 3.06(3) & 3.11(4) \\
\end{tabular}
\end{ruledtabular}
\caption{Length spectrum of finite clusters, power fit parameter}
\label{tbl2}
\end{table*}

\subsection{Monopole densities}
\label{densities}

As we discussed in the introduction the magnetic monopole clusters fall into
two different classes~\cite{uvir,hart}: "small", or finite clusters which have
finite size in lattice units, and "large" clusters which percolate through the
lattice. It was demonstrated ~\cite{hart} that the percolating  cluster 
is responsible for the string tension. 
If the size of the lattice is large enough, there is a natural
distinction between "small" and "large" clusters, since in each configuration
there exists only one cluster which is much longer than all others and there is
a gap in the cluster length distribution~\cite{hart}. If the lattice size is
not large enough the percolating  cluster decays into one or more large clusters
 plus several monopole currents which wind through the boundaries of the lattice
\cite{mueller}.
\begin{figure}[!ht]
\begin{center}
\includegraphics[width=\columnwidth]{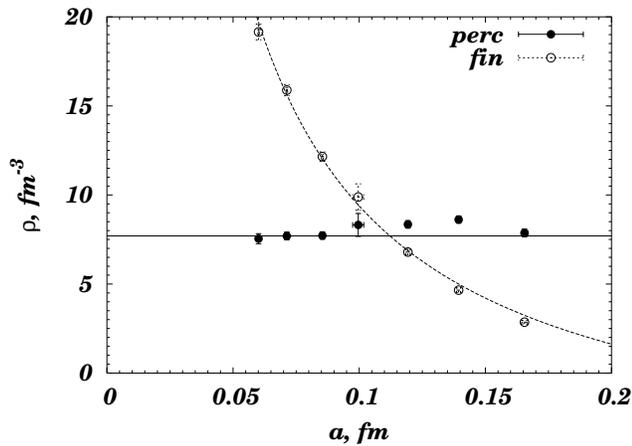}
\caption{Density of the finite clusters $\rho_{fin}$ and percolating clusters
$\rho_{perc}$; solid and dashed 
lines, are fits by a constant  and Eq.(\ref{rhofin})  respectively.}
\label{rho_fig}
\end{center}
\end{figure}
It occurs that the monopole density for
the sum of these large clusters and the winding trajectories scales 
\cite{mueller}. We
simply call this collection of the monopole currents as percolating cluster. 

In Fig.~\ref{rho_fig} we show the monopole density $\rho_{perc}$ as a function
of the lattice spacing $a$. Our results for $\rho_{perc}$ obtained with higher 
statistics than in \cite{mueller} agree within error bars with values
obtained in \cite{mueller} for coinciding values of $\beta$
on $a$ is rather weak and we fit it by a constant for $\beta > 2.35$ 
The resulting density of the percolating
monopoles in the continuum limit is
\begin{equation}\label{rhoperc}
\rho_{perc}~=~7.70(8)\, fm^{-3}~~.
\end{equation}
This value is shown by the solid line in Fig.~\ref{rho_fig}.
It is in agreement with value obtained in \cite{mueller} within error bars.

\begin{figure}
\begin{center}
\includegraphics[width=\columnwidth]{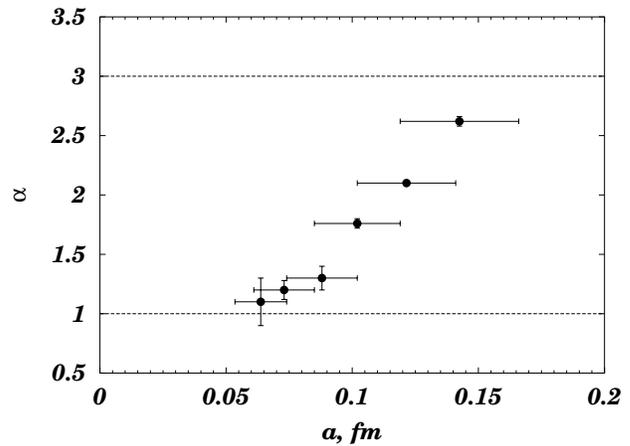}
\caption{The dependence $\alpha(a)$.} \label{alpha}
\end{center}
\end{figure}

The density of the finite clusters, on the other hand, is divergent for $a\to
0$ and it can be fitted (dashed curve on Fig.\ref{rho_fig}) as
\begin{equation}\label{rhofin}
\rho_{fin}~ =~ C_1~ + ~\frac{C_2}{a}\, ,
\label{fit_df}
\end{equation}
where
$$C_1 = -6.1(5)\,fm^{-3}~,~ C_2~ =~ 1.55(4) \, fm^{-2}\, .$$
The negative value of the constant $C_1$ means that the fit is not valid for
large values of the lattice spacing. 

We made also the local fits of $\rho_{fin}$ by function $A\cdot a^{- \alpha }$,
where $A$ and $\alpha$ are fit parameters. The fit results for $\alpha$ 
are shown in Fig.~\ref{alpha}. To get a point on this figure used for the fit
three neighbor data points in Fig.~\ref{rho_fig}. Corresponding fit intervals
are depicted in Fig.~\ref{alpha} as X axes error bars. It can be seen from 
Fig.~\ref{alpha} that the values of $\alpha$ are changing from the value
close to 3 down to the value approximately equal 1, i.e. at small $a$ this 
fit becomes consistent with the fit Eq.(\ref{fit_df}).  

Another possible fit,
suggested in Ref.~\cite{hart}, $\rho_{fin} = A\cdot a^{- \alpha }$ gives $A$
and $\alpha$ strongly dependent on $a$, see Fig.~\ref{alpha}. However, at
smallest $a$ we have the same $1/a$ behavior of the density.

\subsection{Correlator of the monopole currents}
\label{sub33}

We turn now to discussion of correlation functions of the monopole currents.
We will consider
three definitions:

\begin{itemize}

\item $G_1(x-y)$ is the probability that point $x$ and point $y$
are connected
by a monopole trajectory.

\item $G_2(x-y)$ is the probability that point $x$ and point $y$ are connected
by a monopole line belonging to the percolating cluster.

\item $G_3(x-y)$ is the probability that the monopole
current crosses point $x$
and point $y$.

\end{itemize}

Since the small clusters have finite size
in the lattice units (see
Sect.~\ref{short}) the functions $G_1(x-y)$ and $G_2(x-y)$ coincide
with each other for large
$d = |x-y|$. The asymptotic behavior for $d=|x-y| \to
\infty$ is similar for all correlation functions and was suggested first
in \cite{IvPoPo} (see also \cite{reinhardt}):
\begin{equation}\label{masses}
G_k(x-y) \to C_k + A_k \exp\left\{- m_k d\right\} \, , \label{Gasimpt}
\end{equation}
where $k=1,2,3$ and  $m_k$ are expected to be
the mass of the lightest glueball.
Moreover, since by definition (see (\ref{l})) the monopole density
is the probability to find
the monopole current on the given link, the constants $C_k$ are equal to the
squares of the corresponding densities. It is obvious that $C_{1,2} =
\left(
\rho_{perc}\right)^2$ and $C_3 = \left( \rho_{perc} + \rho_{fin}\right)^2$.

\begin{figure}
\begin{center}
\begin{minipage}{0.22\columnwidth}
\begin{center} $d = 1.7a$ \end{center}
\includegraphics[width=1.0\textwidth]{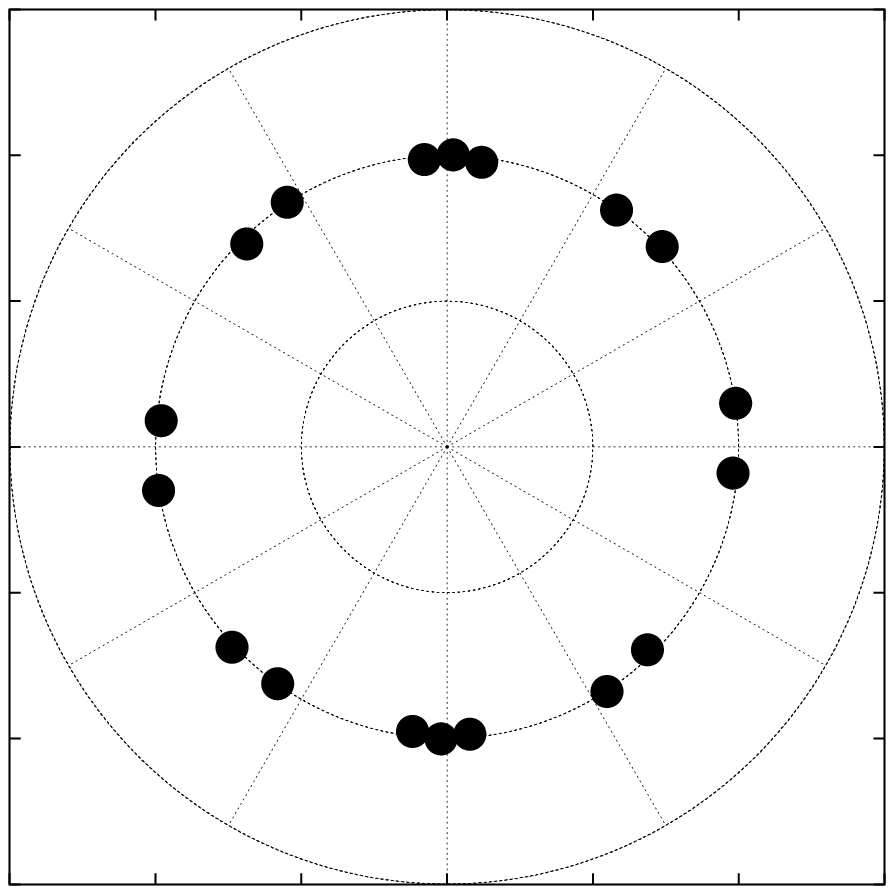} \\
\end{minipage}
\begin{minipage}{0.22\columnwidth}
\begin{center} $d = 2.2a$ \end{center}
\includegraphics[width=1.0\textwidth]{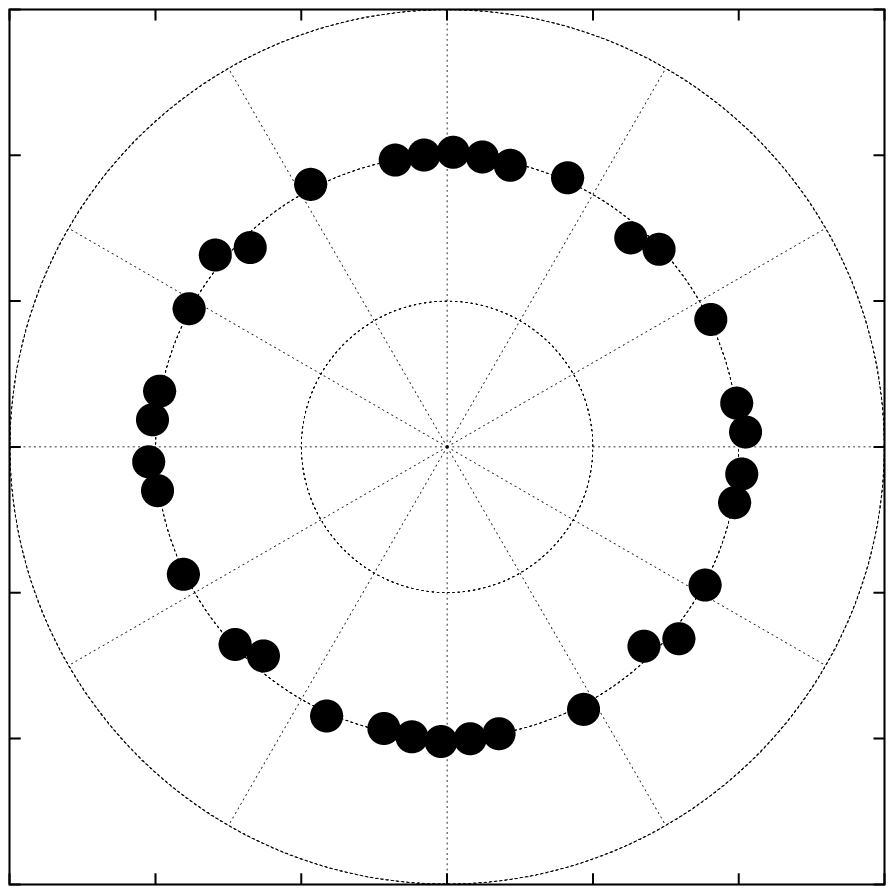} \\
\end{minipage}
\begin{minipage}{0.22\columnwidth}
\begin{center} $d = 3.6a$ \end{center}
\includegraphics[width=1.0\textwidth]{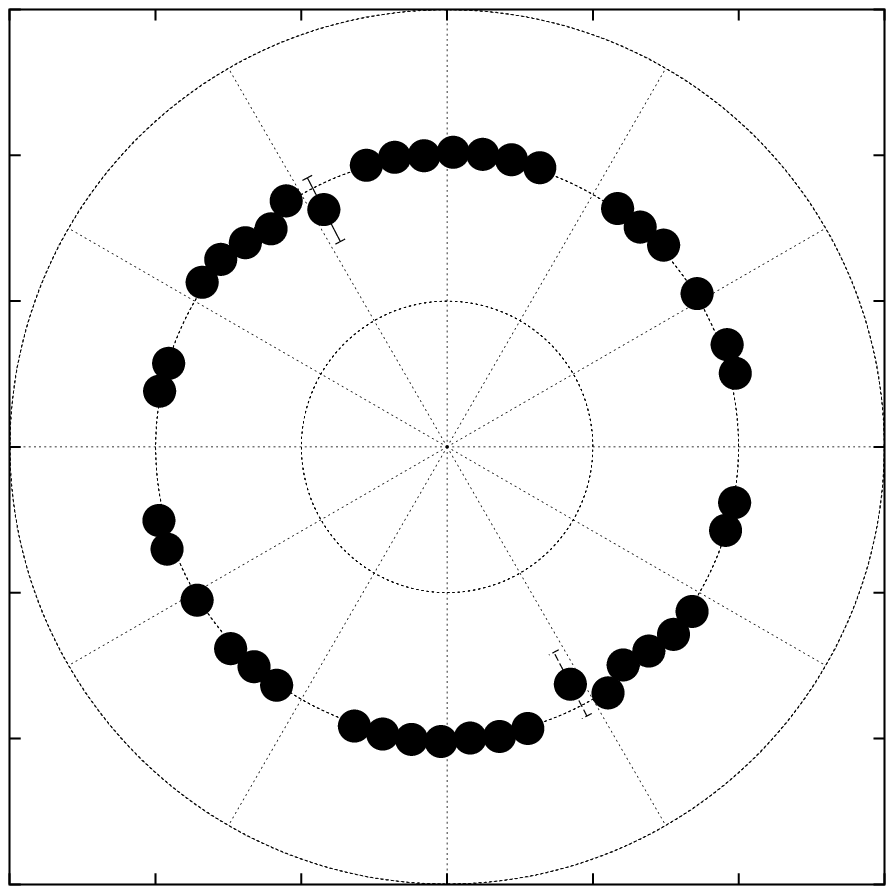} \\
\end{minipage}
\begin{minipage}{0.22\columnwidth}
\begin{center} $d = 9.05a$ \end{center}
\includegraphics[width=1.0\textwidth]{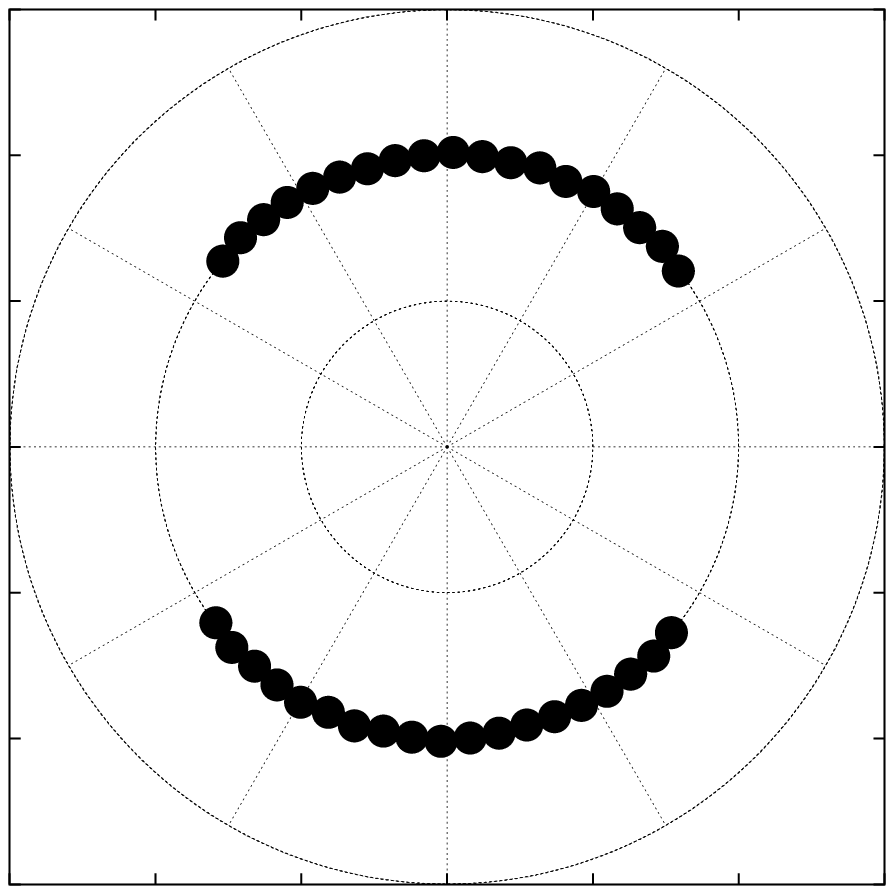} \\
\end{minipage}
\caption {Polar plots for $G_3(d)$ at $\beta = 2.3$, the lattice size is
$16^4$.} \label{rotation}
\end{center}
\end{figure}

\begin{figure}
\begin{center}
\includegraphics[width=\columnwidth]{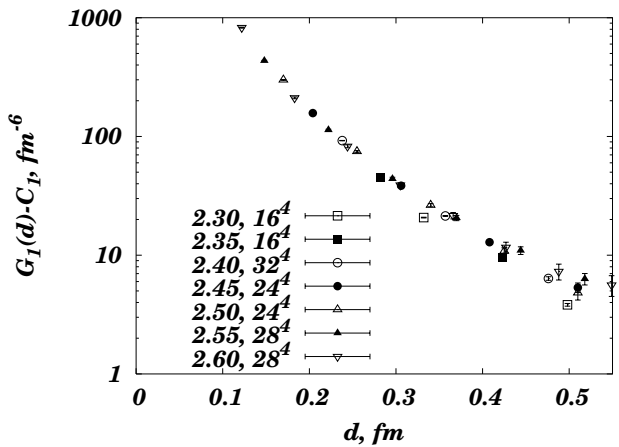}
\caption {The scaling behavior of the correlation function $G_1(d)$.
The constant $C_1$ has been subtracted.}
\label{Gscaling_fig}
\end{center}
\end{figure}

The results of the measurements are presented in Figs.~\ref{rotation},
\ref{Gscaling_fig}. In Fig.~\ref{rotation} we show the results of 
the rotational invariance check 
of the correlators $G_k$ at large enough $d$. Namely,
the dependence of $G_3(x-y)$ on the angle between 4-vector $x-y$ and
one (arbitrary) axis in polar coordinates is shown. 
For illustrative purposes, we use different normalizations of $G_3$ on 
different plots in Fig.~\ref{rotation}.
The data show nice rotational invariance. Note that the gap in the 
angles seen for large $d$ is due to the geometry of the lattice. 

One of the most important properties of the correlators is their scaling. The
data do indicate that the correlators do have the scaling behavior. This is
illustrated in Fig.~\ref{Gscaling_fig}, where we show $G_1(d)$ in physical
units for various $\beta$. Note that the dimension of $\rho$ is $fm^{-3}$.
Respectively, the dimension of the correlation functions $G_k$ is $fm^{-6}$ and
they are sensitive to possible scaling violations. Comparing the data for 
various volumes at $\beta=2.4$ we found the lattice volume
independence of the correlators $G_k(d)$.

Finally, we determined the masses $m_k$ in Eq.~(\ref{Gasimpt}). The fits 
made for the distance $d > 0.3$ fm show that
the masses $m_k$ do have scaling behavior, but we do not have the high
precision in their determination:
$$m_{1,2}~ =~ 1.87 (16)\, GeV ,~
m_3~ =~ 2.4 (5) \, GeV~.$$ Due to large errors
we can claim only that the results do not
contradict the prediction that $m_k$ coincide with the
mass of the lightest $SU(2)$ $0^{++}$ glueball
as measured, in particular, in Refs.~\cite{glueball}.

\section{Percolation laws at short and large distances}
\label{discussion}

In  preceding sections we presented data on the phenomenology of
the monopole clusters and have chosen to postpone  comments on their
implications till this section.
It might worth emphasizing from the very beginning
that not every piece of information that we got has a straightforward
interpretation. Moreover,
it is useful to distinguish between the properties of
the clusters at relatively short
 distances, $\Lambda_{QCD}^{-1}~\gg~l\gg a$ where $l$ is the length
along the monopole trajectory, and at large distances $l~\geq~
\Lambda_{QCD}^{-1}$. As we shall see, it is the short distance behavior which
is quite well understood while for larger distances one could rather speak of
accumulation of the data. This might look paradoxical since traditionally the
non-perturbative fluctuations are deprived of interesting short-distance
behavior.

\subsection{ Monopoles as free particles at short distances}

One of our main results is observation of scaling
for a few observables :

\begin{itemize}

\item average distance between the self-crossings
of the percolating cluster, as measured along the trajectory, scales
(see Fig. \ref{lscaling_fig});

\item average Euclidean distance between the same points approximately
scales, see Fig. \ref{dscaling_fig};

\item correlation length of direction alignment of links along the trajectory
of the percolating cluster scales, see Table~\ref{mu};

\item correlators of the vacuum monopole currents scale, as illustrated in
Fig.~\ref{Gscaling_fig};

\item we confirm scaling of the density
of the percolating cluster, see Fig.~\ref{rho_fig}.

\end{itemize}

Each of these checks is highly non-trivial since the most natural explanation
of any scaling behavior is that the monopoles of size $a$ are physically
meaningful. Thus, each example of the scaling behavior confirms, at least
indirectly, fine tuning of the lattice monopoles.

Now, let us discuss briefly the geometry of the monopole cluster
at various distances. At the very short distances, of order $a$,
monopole trajectories can be thought of as (approximate) random walk.
At these distances both the action and the entropy are of order $1/a$.
Let us illustrate this statement by a well known equation
(see, e.g., \cite{ambjorn}) for the propagating
mass of a scalar particle $m^2_{prop}$:
\beq\label{propagator}
m^2_{prop}\cdot a~\approx~\big(M(a)~-~{\ln 7\over a}\big)~~,
\eeq
$M(a)$ is the field theoretical mass of the same particle,
discussed in the Introduction on the monopole example.
A more general definition of $M(a)$ is that the
action of the particle is $$
S_{cl}~=~M(a)\cdot l~~,$$
where $l$ is the length of a path connecting two points between which
the particle propagates. Moreover, the $\ln 7$ factor
is due to the entropy. Eq. (\ref{propagator}) is derived in the approximation
of neglecting neighboring trajectories. Note a dramatic difference
between $M(a)$ and $m_{prop}$. Namely, to ensure a finite propagating mass
$m_{prop}$ one needs an ultraviolet divergent mass $M(a)$
(that is, the mass which determines the
classical action).

Upon accounting for possible cancellations between the action and
entropy one can introduce an `effective action',
$$S_{eff}~=~l\cdot \mu~,~~(l~\gg ~a)~,$$
which governs the behavior of the trajectories at distances much larger than
$a$. The very fact that the density of the percolating cluster scales implies
that at distances of order $a$ the divergent factors in the action and entropy
cancel each other and $\mu~\sim~\Lambda_{QCD}$. This is just the fine tuning
hypothesis which we heavily rely on.

Eq. (\ref{propagator}) is useful for orientation in mass scales which came out
from various fits to the data. In particular fits to the correlation of the
links resulted in the attenuation factor of order $\exp(-\mu_sl)$ where $\mu_s$
is about $300~ MeV $ and $l$ is measured along the trajectory, see Table
\ref{mu}. At first sight, $300~MeV$ might look too low a mass. However, if we
apply for estimates Eq. (\ref{propagator}) we would obtain rather
$m^2_{prop}~\sim~(300~MeV)/a$. In other words, we do have evidence that the
most singular part of the action (as measured on the lattice, see the
Introduction) is indeed cancelled by the entropy factor. But the value
$\mu_s~\approx~ 300~MeV$ does not necessarily corresponds directly to a
physical mass.

Once this cancellation is confirmed by the data, one can make predictions
about short clusters with the length $\Lambda_{QCD}^{-1}~\gg~l~\gg~a$
\cite{maxim}. Namely, at such length one can neglect the mass factor
and the spectrum of the closed loops should correspond to free
particles. Then one can derive (see \cite{maxim} and references therein):
\beq\label{free}
N(l)~\sim~{const\over l^3}~,~~~R(l)~\sim\sqrt{l}~~,
\eeq
where $R(l)$ is the radius of the cluster of the length $l$.
Let us emphasize that (\ref{free}) is a consequence
of free field theory in Euclidean space-time
(actually, in $d=4$ Eq. (\ref{free})
survives adding Coulomb like interaction as well \cite{maxim}).
Eq. (\ref{free}) is derived most naturally in the so called polymer
representation, see, e.g., \cite{polyakov1,maxim}.

Measurements of $N(l)$ and $R(l)$ were reported in \cite{hart} and are in
perfect agreement with (\ref{free}). Our analysis confirms (\ref{free}) on
larger statistics and for smaller values of $a$. In particular, the data on the
$l$ distribution are summarized in the Table~\ref{tbl2}. We did not include the
data on the $R(l)$ which we have but they do confirm (\ref{free}) for all $a$
tested.

\subsection{Percolating cluster}

Existence of a single infinite cluster is typical for all percolating systems
in the supercritical phase, see, e.g., \cite{grimmelt}. There is a general
theorem on the uniqueness of the percolating cluster at $p~>~p_c$ where $p$ is
the probability to have a link `open' (in our case, to belong to a monopole
trajectory) and $p_c$ is the point of the phase transition. The percolating
cluster is characterized by the probability $\theta(p)$ of a given link to
belong to the cluster (in our case this probability is equal to the cluster
density, $\rho_{perc}$). Generically, $\theta(p)\sim(p-p_c)^{\beta}\,
,~\beta>0$. In our case, \beq \label{theta}
\theta(p)~\sim~(a\cdot\Lambda_{QCD})^3~\ll~1~.
\eeq
Smallness $\theta(p)$
implies closeness to the point of phase transition to the percolation.
In the limit $a\to 0$ we hit exactly the point of phase transition.
Expression (\ref{theta}) can be considered as formulation of the
hypothesis on fine tuning of the monopoles in terms of the percolation
theory.

Another general property of the percolating cluster is that its
fractal dimension coincides with the dimension of the space:
\beq
D_{fr}^{perc}~=~d~~.
\eeq
Which is trivially satisfied in our case, since the length of the percolating
cluster is proportional to the whole volume, $V_4$.

However, to the best of our knowledge, the percolation
theory is not powerful enough to  predict  or explain
the data which we have on such characteristics as
$l_{segm},d$.

It is worth emphasizing that geometrical characteristics of the percolating
cluster at length of order $\langle l_{segm} \rangle$ differ quite drastically from
the corresponding characteristics of the finite clusters.
First of all, the monopole trajectories are no longer random walks.
Indeed, for a random walk we would have had:
\beq\label{randomwalk}
l~\approx~{1\over 2}{d\over m_{prop}\cdot a}~~,
\eeq
where $l$ is the distance between two points measured
along the trajectory and $d$ is Euclidean distance between the same points.

Our results on $l_{segm}$ and $d$ are given in Fig. \ref{lscaling_fig} and
\ref{dscaling_fig}, respectively. The data certainly rule out
(\ref{randomwalk}) for any finite $m_{prop}$. Moreover if we assume that it is
only the leading power of $1/a$ which is cancelled in Eq. (\ref{propagator})
then $m_{prop}\sim \sqrt{\Lambda_{QCD}/a}$. Our data are not consistent with
(\ref{randomwalk}) for such mass either.

Also, existence of correlation between directions of the links, (see Sect.
\ref{linksdirection}), could not be reconciled with the random walk.

\subsection{Long-range correlation of finite clusters}
\label{finclust}

Turn now to discussion of the data on the density of the short clusters,
see Sect. \ref{densities}. Expectations for $\rho_{fin}$ are easy to formulate.
Indeed, the spectrum $N(l)\sim l^{-3}$ implies that the average length
of the finite clusters is saturated in the ultraviolet, i.e. at length
of order $a$. Then the density of the finite clusters should not depend
on $\Lambda_{QCD}$ at all and on the dimensional grounds
$\rho_{fin}\sim a^{-3}$.

Instead, we find that the most singular piece in the density $\rho_{fin} \sim
\Lambda_{QCD}^2\cdot a^{-1}$. The only interpretation of this striking
observation is that even short, or finite clusters are not independent at large
scales. Again, on dimensional grounds alone it is clear that the finite
clusters are in fact associated with two-dimensional surfaces whose area
scales.

At first sight, this conclusion looks too bizarre. However,
in fact it is known to be true from independent measurements
on the P-vortices. Namely, it was observed first in Ref.  \cite{giedt}
that monopoles are associated predominantly with P-vortices, whose
total area scales. This conclusion was recently reinforced by measurements
for smaller lattice spacings \cite{kovalenko}.

Thus, we are invited to think about percolation of monopoles as a two-step
process: monopoles percolate on the surface of dimension $d=2$ and the surface
itself percolates in $d=4$ space-time. As far as we know, this type of
percolation has not been studied theoretically at all. And we can, therefore,
suggest only very preliminary considerations.

First the probability $\theta(p)$, see Eq. (\ref{theta}) is to be thought
of as a product of two factors:
\beq\label{split}
\theta(p)~\sim ~(a\cdot\Lambda_{QCD})^3~\sim ~(a\cdot \Lambda_{QCD})
\cdot {V_2\over V_4}~~,
\eeq
where the first factor is to be interpreted as a power of $(p^{(2)}-p^{(2)}_c)$
for percolation of monopoles on a surface while the second factor accounts for
suppression of the `phase space' available on a $d=2$ space, $V_2$, as compared
to the whole lattice volume $V_4$. Note that for probabilities defined on the
surface we still have according to (\ref{split})
$$(p^{(2)}-p^{(2)}_c)~\sim ~(a\cdot \Lambda_{QCD})~\ll~1, ~~a\to 0~.$$

Another major change, is that finite clusters now occupy a finite fraction of
the links on the surface. Furthermore, one could also speculate that for finite
monopole clusters of length $l\sim \Lambda_{QCD}^{-1}$ we would still have
$$R~\sim~ l^{1/d} ~\sim~ \sqrt{l}~~,~ d=2$$

As the last remark, let us notice that the interplay between the monopole
trajectories and P-vortices is more complicated than simply saying `monopoles
belong to P-vortices'. Indeed, for short clusters the length spectrum is
sensitive to the dimension of the space~\cite{polyakov1,maxim}:

\beq N(l)~\sim~{1\over l^{d/2+1}}~~,
\eeq
and the spectrum $\sim l^{-3}$ corresponds to $d=4$. In this sense, at short
distances the monopole trajectories are `primary' objects. Thus, one can
expect that for larger $l$ where the infrared behavior of the clusters sets in
the spectrum is changed into $1/l^2$.

\section{Conclusions}

There is accumulated evidence that monopoles defined within the Maximal
Abelian projection appear physical even if studied with resolution of
the lattice spacing $a$. In particular, we have seen that there are
amusingly simple scaling laws for various observables which depend
only on the product $(a\cdot \Lambda_{QCD})$ and look perfectly
gauge-invariant. Thus, non-perturbative fluctuations in the lattice
$SU(2)$ appear to have non-trivial structure in the ultraviolet
as well as in the infrared.

What is specific for the field, is that phenomenology seems to be
far ahead of theory. In particular, we have argued that there are
indications that percolation of the monopoles is to be considered
as combination of percolation of the trajectories over a surface $d=2$
and of percolation of the surface over the $d=4$ space.
To the best of our knowledge, no theoretical framework is available
to describe such a percolation.

\section*{Acknowledgements}
The authors are thankful to M.~M\"uller-Preussker for contribution
at the initial stage of this work. We are obliged to
F.V.~Gubarev, M.N.~Chernodub,  A.V.~Kovalenko and S.N.~Syritsyn for numerous
fruitful discussions. V.~G.~B., P.~Yu.~B. and M.~I.~P. are partially supported
by grants RFBR 02-02-17308, RFBR 01-02-17456, DFG-RFBR 436 RUS 113/739/0,
INTAS-00-00111 and CRDF award RPI-2364-MO-02.

\begin{table*}
\begin{center}
\begin{ruledtabular}
\begin{tabular}{c l l l l l l l l l l}
$\beta$ & 2.30 & 2.35 & 2.40 & 2.40 & 2.40 & 2.40 & 2.45 & 2.50 & 2.55 & 2.60 \\
$a\, (fm)$ & 0.1655(13) & 0.1394(8) & 0.1193(9) & 0.1193(9) & 0.1193(9) & 0.1193(9) & 0.0996(22) & 0.0854(4) & 0.0713(3) & 0.0601(3) \\
$L$ & 16 & 16 & 16 & 24 & 28 & 32 & 24 & 24 & 28 & 28 \\
{\verb"#"} of conf. & 100 & 100 & 300 & 137 & 14 & 35 & 20 & 50 & 40 & 50 \\
\end{tabular}
\end{ruledtabular}
\caption{SU(2) configurations} \label{config}
\end{center}
\end{table*}
\section*{Appendix}
The list of configurations used in this work is given in Table~\ref{config}.
To fix the Maximal Abelian gauge we use the Simulating Annealing
algorithm~\cite{SA} and study 10 randomly generated gauge copies for each
configuration to avoid the Gribov copy problem.

Most of the quantities in the paper are given in physical units ({\em e.g.} in
{\it fm}). In order to express the lattice data in physical units we use the
data of Refs.~\cite{physunit} for the $SU(2)$ string tension and assume that
$\sqrt{\sigma} = 440\, MeV$. The corresponding values of the lattice spacing
are given in Table~\ref{config}.

\newcommand{\plb}[1]{{\it Phys. Lett.} {\bf B#1}\ }
\newcommand{\npb}[1]{{\it Nucl. Phys.} {\bf B#1}\ }
\newcommand{\prdd}[1]{{\it Phys. Rev.} {\bf D#1}\ }


\begin{thebibliography}{99}
\bibitem{reviews}
M.N. Chernodub, M.I. Polikarpov, in {\it 'Cambridge 1997, Confinement, duality,
and nonperturbative aspects of QCD'}, p.~387; {\tt hep-th/9710205}.

\bibitem{mueller}
V.~Bornyakov and M.~M\"uller-Preussker,
{\it Nucl.\ Phys.} {\bf B106} (2002) 646.

\bibitem{hart} A.~Hart and M.~Teper, \prdd{58} (1998) 014504;\\
\prdd{60} (1999) 114506.

\bibitem{coleman}
S. Coleman, in {\it ``The Unity of the Fundamental Interactions''},
Erice lectures, 1981, ed. A. Zichichi, Plenum, London (1983) p. 21;\\
M.N. Chernodub, F.V. Gubarev, M.I. Polikarpov, V.I. Zakharov,
{\it Phys. Atom. Nucl.} {\bf 64} (2001) 561, {\it Yad. Fiz.}
{\bf 64} (2001) 615, ({\tt hep-th/0007135}).

\bibitem{anatomy}
B.L.G.~Bakker, M.N.~Chernodub and  M.I.~Polikarpov, {\it Phys. Rev. Lett.} {\bf
80} (1998) 30;\\
V.G.~Bornyakov, F.V.~Gubarev, M.I.~Polikarpov, T.~Suzuki, A.I.~Veselov and
V.I.~Zakharov, {\it Phys. Lett.} {\bf B537} (2002) 291.

\bibitem{polyakov}
A.M. Polyakov, {\it Phys. Lett.} {\bf B59} (1975) 82;\\
T. Banks, R. Meerson, J. Kogut, {\it Nucl. Phys.} {\bf B129} (1977) 493.

\bibitem{vz}
V.I. Zakharov, {\it ``Hidden mass hierarchy in QCD''}, {\tt hep-ph/020404}.
\bibitem{maxim}
M.N. Chernodub, V.I. Zakharov, {\it ``Towards understanding structure of the
 monopoles clusters''}, {\tt hep-th/0211267}.


\bibitem{latt02} P.Yu.~Boyko, M.I.~Polikarpov and V.I.~Zakharov,
{\it ``Geometry of percolating monopole clusters''},
{\tt hep-lat/0209075}.

\bibitem{uvir} S.~Kitahara, Y.~Matsubara and T.~Suzuki, 
{\it Progr. Theor. Phys.} {\bf 93}, 1 (1995);\\
M.~Fukushima, A.~Tanaka, S.~Sasaki, H.~Suganuma, H.~Toki and
D.~Diakonov, {\it Nucl.
Phys. Proc. Suppl.} {\bf 53}, 494 (1997).


\bibitem{IvPoPo}
T.~L.~Ivanenko, M.~I.~Polikarpov and A.~V.~Pochinsky, {\it JETP Lett.}\  {\bf 53}
(1991) 543 [{\it Pisma Zh.\ Eksp.\ Teor.\ Fiz.\ } {\bf 53} (1991) 517];\\
T.~L.~Ivanenko, A.~V.~Pochinsky and M.~I.~Polikarpov,
{\it Phys.\ Lett.} {\bf B302} (1993) 458.

\bibitem{reinhardt}
K. Langfeld, H. Reinhardt, {\it ``Monopole~-~ anti-monopole
excitation in MAG projected SU(2) lattice gauge theory''},  {\tt hep-lat/0206021}.

\bibitem{glueball}
C.~Michael and M.~Teper, {\it Nucl. Phys.} {\bf B305} (1988) 453;\\
C.~Michael and S.~Perantonis, {\it J. Phys.} {\bf G18} (1992) 1725;\\
S.~Booth et al (UKQCD), {\it Nucl. Phys.} {\bf B394} (1993) 509.

\bibitem{ambjorn}
J. Ambjorn, {\it ``Quantization of geometry''}, {\tt hep-th/9411179}.

\bibitem{polyakov1}
A.M. Polyakov, {\it ``Gauge Fields and Strings''}, Harwood
academic publishers, (1987).

\bibitem{grimmelt}
C. Itzykson, J.-M. Drouffe, {\it ``Statistical Field Theory''},
vol. 1, Cambridge University Press, (1989);\\
D. Stauffer, A. Aharony,
  {\it  ``Introduction to percolation theory''}
   London et al.: Taylor and Francis 1994;\\
G. Grimmett,
  {\it  ``Percolation''}
   Berlin et al.: Springer 1999,
   Grundlehren der mathematischen Wissenschaften, Vol. 321.

\bibitem{giedt}
J. Ambjorn, J. Giedt, J. Greensite, {\it JHEP} {\bf 9903}
(1999) 019, {\tt hep-lat/9903023}.

\bibitem{kovalenko}
F.V. Gubarev, A.V. Kovalenko, M.I. Polikarpov, S.N. Syritsyn, V.I. Zakharov,
{\it ``Fine tuned vortices in lattice $SU(2)$ gluodynamics},
 {\tt hep-lat/0212003}.

\bibitem{SA}
G.~S.~Bali, V.~Bornyakov, M.~Muller-Preussker and K.~Schilling,
{\it Phys.\ Rev.} {\bf D54} (1996) 2863.

\bibitem{physunit}
J.~Fingberg, U.~M.~Heller and F.~Karsch, {\it Nucl.\ Phys.\ B} {\bf 392} (1993) 493. {\tt hep-lat/920812}\\
G.~S.~Bali et.al. {\it Int.J.Mod.Phys.} {\bf C4} (1993) 1179-1193 
{\tt hep-lat/9308003} \\
G.~S.~Bali, K.~Schilling, A.~Wachter {\it Phys.Rev. D} {\bf 55} (1997) 5309-5324 {\tt hep-lat/9611025} \\
G.~S.~Bali, K.~Schilling, C.~Schlichter {\it Phys.Rev. D} {\bf 51} (1995) 5165-5198 {\tt hep-lat/9409005} \\
B.~Lucini, M.~Teper {JHEP} {0106} (2001) 050 {\tt hep-lat/0103027} 

\end{thebibliography}
\end{document}